\documentclass[journal=nalefd,manuscript=letter]{achemso}
\usepackage[version=3]{mhchem} 
\usepackage{xcolor}

\author{Andreas Ørsted}
\email{andreas.orsted@unige.ch}
\author{Alessandro Scarfato}
\author{Céline Barreteau}
\author{Enrico Giannini}
\author{Christoph Renner}
\email{christoph.renner@unige.ch}
\affiliation{University of Geneva, Department of Quantum Matter Physics, 24 Quai Ernest-Ansermet 1211 Geneva 4, Switzerland}
\title{Doping Tunable CDW Phase Transition in Bulk 1T-ZrSe$_2$}

\begin{document}

\begin{abstract}
Tuneable electronic properties in transition metal dichalcogenides (TMDs) are essential to further their use in device applications. Here, we present a comprehensive scanning tunnelling microscopy and spectroscopy study of a doping-induced charge density wave (CDW) in semiconducting bulk 1T-ZrSe$_2$. We find that atomic impurities which locally shift the Fermi level ($E_F$) into the conduction band trigger a CDW reconstruction concomitantly to the opening of a gap at $E_F$. Our findings shed new light on earlier photoemission spectroscopy and theoretical studies of bulk 1T-ZrSe$_2$, and provide a local understanding of the electron-doping mediated CDW transition observed in semiconducting TMDs.
\end{abstract}

\maketitle
Since the discovery of graphene, extensive theoretical and experimental efforts have been devoted to two-dimensional (2D) and quasi-2D systems. The investigations into their intriguing physical properties have generated numerous highly influential contributions to the field of material physics, including 2D superconductivity\cite{saito2016highly}, heterostructure devices\cite{novoselov20162d}, moiré physics\cite{he2021moire}, topology\cite{kou2017two}, spin and charge density waves\cite{Hwang2024}, and many more. However, even though our understanding of these systems has grown massively in recent years, with both theoretical and experimental advances, individual pieces of the puzzle remain undiscovered or elusive. Here, we focus on the charge density wave (CDW) phase in a quasi-2D system, which is of particular interest because of the many open questions concerning its formation mechanism and its interplay with other electronic phases such as superconductivity\cite{morosan2006superconductivity,chang2012direct,singh2005competition}.

The CDW formation in one dimension is well-established: a metallic chain of atoms distorts into a gapped state concomitant to a modulated charge distribution due to Fermi surface nesting (FSN). However, the underlying mechanism of the CDW phase transition in 2D and quasi-2D compounds remains to be fully understood. In certain systems, FSN is proposed as the driving force\cite{borisenko2009two,bosak2021evidence,strocov2012three,giambattista1990scanning}, while in others, Jahn-Teller-like electron-phonon coupling is considered the primary candidate\cite{johannes2008fermi,zhu2015classification,valla2004quasiparticle,xie2022electron,weber2011electron,otto2021mechanisms,si2020origin}. 

We use scanning tunnelling microscopy (STM) and spectroscopy (STS) in ultra-high vacuum to investigate in-situ cleaved surfaces of bulk 1T-ZrSe$_2$ (hereafter ZrSe$_2$) at 4.5~Kelvin. Depending on scanning bias voltage, we find dispersing and non-dispersing periodic charge modulations. At small negative tunnelling biases, we observe non-dispersing charge modulations akin to the $2a\times 2a$ modulation observed at the semiconductor-to-metal transition in few-layer thin ZrSe$_2$ flakes grown by molecular beam epitaxy on graphitized SiC(0001)\cite{ren2022semiconductor}. A well-defined non-dispersing \textbf{\textit{q}}-vector and contrast inversion across the CDW gap at the Fermi level ($E_F$) underline the CDW nature of this modulation\cite{spera2020}. At a tunneling bias of $+100$~meV or higher above $E_F$, we find dispersive charge modulations with a wavelength that increases with energy. These modulations are clearly quasiparticle interference (QPI) patterns incompatible with a CDW. They are associated with a large density of state (DOS) in the conduction band and only observed above $E_F$.

The scanning probe data discussed here provide compelling evidence that the onset of the CDW phase transition in bulk ZrSe$_2$ is associated with a shift of the Fermi level into the conduction band due to local electron doping by impurities. A doping-induced CDW concomitant to a semiconductor-to-metal phase transition has previously been reported in few-layers thin ZrSe$_2$ grown on graphene.\cite{ren2022semiconductor} However, in contrast to these previous experiments, we can unambiguously attribute the CDW's origin to doping, excluding any measurable strain or reduced dimensionality. Our data reveal a remarkable correlation between the shift of the Fermi level into the conduction band and the appearance of a CDW with the opening of a gap at $E_F$. ZrSe$_2$ is the second system after potassium doped MoS$_2$ where a doping-dependent CDW reconstruction is observed\cite{bin2021charge}, demonstrating doping as a possible tuning parameter of the CDW ground state in semiconductor TMD systems. 

Single crystals of 1T-ZrSe$_2$ were grown by the chemical vapour transport (CVT) method, using ZrCl$_4$ as a transport agent. The use of ZrCl$_4$ is beneficial for the growth of high-quality crystals, compared to the more common use of iodine as a transport agent. Adding low amounts of the transition metal chloride to the precursors, instead of pure iodine, was reported to favour the fast and reproducible growth of pure crystals of many other TMDs \cite{Ubaldini2013TMchlorides}. We started from 99.9\%-pure Zr lumps, 99.5\%-pure ZrCl$_4$ powder and 99.999\%-pure Se shots, according to the following reaction equation 0.9\,Zr + 0.1\,ZrCl$_4$ + 2\,Se $\longrightarrow$ ZrSe$_2$ + 0.2\,Cl$_2$. The precursors were weighed and mixed inside a glove box under Ar(6N) atmosphere, then sealed under vacuum in a quartz tube of internal diameter 8\,mm and length $\simeq$12\,cm. A total mass of 0.3\,g was introduced into each ampule. The ampule was annealed in a tubular furnace under a T-gradient (T$_{hot}$=830$^\circ$C, T$_{cold}$=750$^\circ$C) for 120\,h, then quickly pulled out to room temperature. An almost complete transport of the material from the hot to the cold end was achieved under these conditions, and grey-greenish platelets with a metallic lustre were found to crystallise at the cold end.
X-ray diffraction measurements, performed in a Philips X'Pert four-circle diffractometer using Cu$K_\alpha$ radiation, confirmed the 1T-ZrSe$_2$ crystal structure ($P\Bar{3}m_1$ space group, CdI$_2$ structure type) and the [00$l$] orientation of the plate-like crystals. Careful analyses of the crystal composition were carried out by Energy Dispersive X-ray Spectroscopy (EDS) in a LEO438VTP scanning electron microscope coupled to a Noran Pioneer X-ray detector. All crystals exhibited a little excess of Zr and Se vacancies (average composition Zr$_{1.08}$Se$_{1.92}$), locally varying from ZrSe$_2$ to Zr$_{1.15}$Se$_{1.85}$, in agreement with previous reports \cite{Whitehouse1978ZrSe2}. No traces of Cl and ZrSe$_3$ were found within the resolution of the EDS probe. The intrinsic and non-homogeneous off-stoichiometry of ZrSe$_2$ is of primary importance in understanding the electronic properties of this material. The equilibrium Zr-Se phase diagram has not been assessed; however, it is reasonable to expect that it exhibits a broad miscibility range like the similar systems Ti-Se, Zr-S and Zr-Te, and we expect some level of self-intercalated Zr. The crystals were cleaved in-situ shortly before inserting into the STM head.

We carried out STM and STS measurements using etched tungsten tips in a SPECS JT-STM setup operated at $\approx4.4$~Kelvin at a base pressure around $4\times10^{-11}$~mbar. STM measurements were performed in constant current mode, applying a bias voltage to the sample. We have added a small windowing to smooth the edges of all images, resulting in cleaner Fourier transforms. STS conductance maps were acquired using a standard lock-in technique with a $5$~mV AC modulation. CITS maps were taken on a $170 \times 170$ grid in a $30 \times 30$~nm$^2$ window. Each spectrum consists of 183 equally spaced points between $-300$~mV and $+50$~mV and is low-pass filtered using a moving average over five points. The CITS maps are normalized by dividing the $dI/dV(V)$ signal by $(I/V)$. The Fourier components were isolated using soft 21-by-21-pixel Gaussian windows centred on the Fourier peaks.

\begin{figure*}[!htbp]
\centering
  \includegraphics[width=1\linewidth]{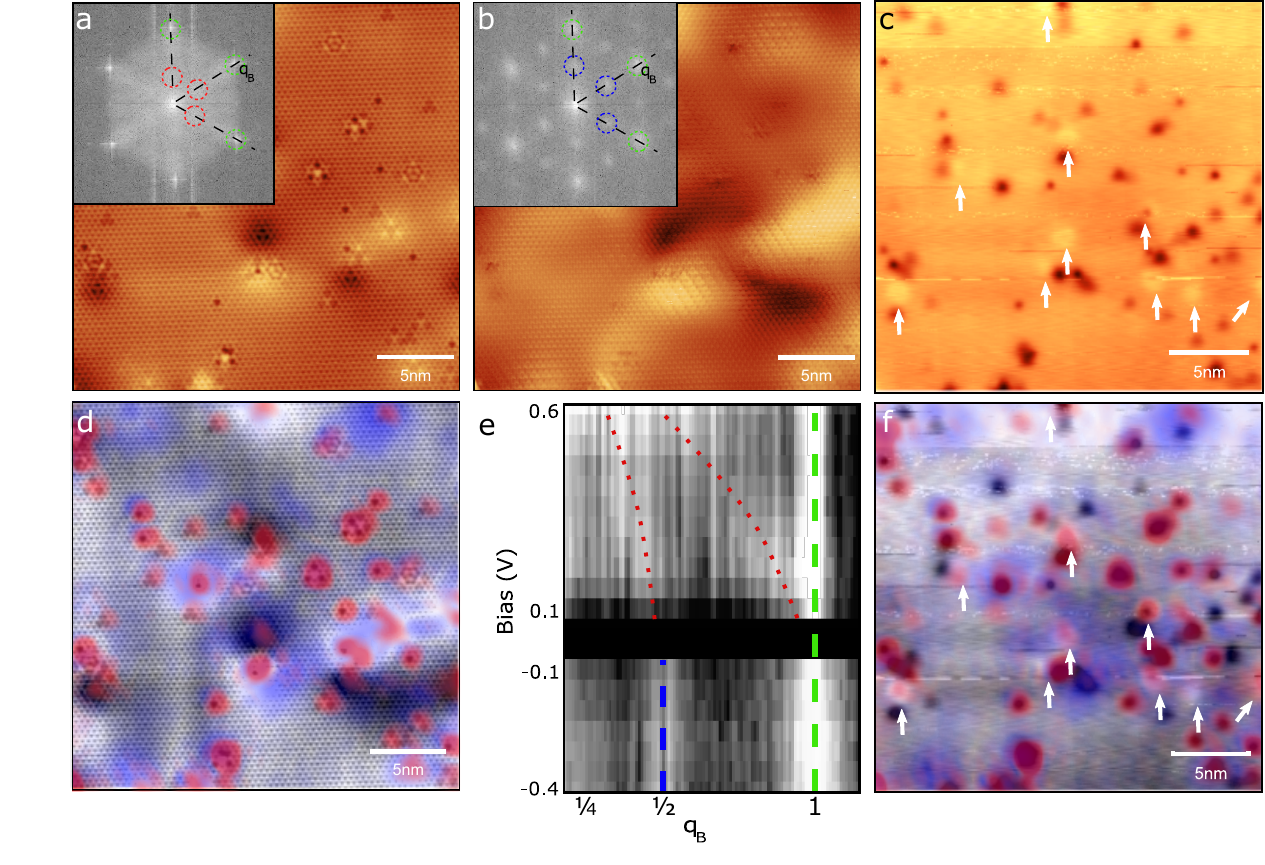}  
\caption{$30\times 30$ nm$^2$ STM topography of the same region of ZrSe$_2$ measured at (a) $+400$~mV, (b) $-400$~mV and (c) $-1.1$~V bias. The insets show the corresponding Fourier transforms. (d) Topography in (a) with overlaid inverse Fourier transforms restricted to the blue and red areas of the FTs in (a) and (b). 
(e) Trace from the $\Gamma$--point to the Bragg peaks as a function of energy averaged along the dashed lines in the FTs (no data is available near $E_F$). The green dashed line marks the Bragg peak, the blue one marks the CDW peak, and the red dotted lines mark two dispersing QPI features. (f) Topography in (c) with overlaid inverse Fourier transforms restricted to the blue and red areas of the FTs in (a) and (b). The white arrows indicate selected hole-doping defects.}
\label{fig: Collected123}
\end{figure*}

In Figure~\ref{fig: Collected123}, we present three constant-current STM topography images acquired at different biases over the same area of an \textit{in-situ} cleaved ZrSe$_2$ surface at 4.4~Kelvin. Positive bias images reveal very sharp atomic resolution with different atomic impurity features (Figure~\ref{fig: Collected123}a). Reducing the bias to small negative voltages (Figure~\ref{fig: Collected123}b), we still resolve the atomic lattice, albeit less sharply, but we no longer see most of the atomic defects resolved at positive bias. The most remarkable aspect of small negative bias images is the appearance of static $2a$ charge modulations aligned with the atomic lattice, where $a$ is the atomic lattice constant. These modulations are only resolved for bias voltages between $-800$~mV and $+50$~mV. At larger negative bias ($<-900$~mV), STM images reveal nm-sized bright and dark spots on a homogeneous background, with no atomic-scale resolution (Figure~\ref{fig: Collected123}c). We associate these spots with the charge inhomogeneities observed at small negative bias in Figure~\ref{fig: Collected123}b, but with opposite contrast. This change, in contrast, is a direct consequence of the semiconducting gap~\cite{LeQuang_2018}, as explained below.

The periodic modulations observed in the topographic and spectroscopic images can be easily isolated by Fourier transform (FT), with distinct features observed in the positive and negative bias images of ZrSe$_2$. The crisp atomic lattice resolved at positive bias yields well-defined Bragg peaks at $q_B$ outlined in green in the inset of Figure~\ref{fig: Collected123}a. At negative bias, the Bragg peaks are weaker and significantly more diffuse (inset in Figure~\ref{fig: Collected123}b). The focus of this study is on the Fourier components near $q_{1/2}=\frac{1}{2}q_B$ present at both polarities but with very different bias dependencies (Figure~\ref{fig: Collected123}e). FTs of positive bias images show very weak peaks near $q_{1/2}$ outlined in red in Figure~\ref{fig: Collected123}a. Their centre of mass shifts towards smaller \textbf{\textit{q}}-values with increasing imaging voltage (Figure~\ref{fig: Collected123}e) while becoming more and more diffuse. Higher positive bias images show finite scattering amplitude in the entire hexagonal region defined by the six Bragg peaks. In contrast, the FTs of negative bias images reveal more defined peaks at $q_{1/2}$ outlined in blue in Figure~\ref{fig: Collected123}b. These peaks do not shift as a function of tunnelling bias (Figure~\ref{fig: Collected123}e), and no additional diffuse amplitude covers the hexagonal region defined by the Bragg peaks. 

The spatial distribution and amplitude of the dispersing and non-dispersing modulations introduced above can be obtained through the inverse FT of the components outlined in blue and in red in Figure~\ref{fig: Collected123}a,b. The composite images obtained by overlaying these spatial distributions on the topographic images
in Figure~\ref{fig: Collected123}d,f clearly show that the dispersing (red) and non-dispersing (blue) components originate in different regions on the surface. The dispersing modulations develop in the immediate vicinity of the sharp atomic defects observed in positive bias topographic images and rapidly decay with distance from them. In contrast, the non-dispersing components are primarily located in between them. As demonstrated below, the dispersing charge modulations observed above $+100$~mV are the result of quasiparticle interference, whereas the non-dispersing modulations with periodicity $2a$ observed between $-800$~mV and $+50$~mV are CDWs. The latter appear only in regions where local electron doping shifts the Fermi level into the conduction band. 

The periodic modulations observed near $q_{1/2}$ at positive and negative bias are of a very different nature. The dispersing \textbf{\textit{q}}-vector as a function of tunnelling bias (Figure~\ref{fig: Collected123}e) and the absence of contrast inversion clearly identify the modulations observed at positive bias above $+100$~mV as quasiparticle interference. They are associated with the sharp rise of the tunnelling conductance with bias above the semiconducting gap, corresponding to a large density of states available for scattering. Many bands contribute to this DOS, with many possible scattering vectors which ultimately cover the entire phase space inside the hexagon defined by the Bragg peaks (see inset in Figure~\ref{fig: Collected123}a). Note that the Bragg peaks are very sharp, indicating that they are not affected by higher-order components that would be expected if the peaks near $q_{1/2}$ were a CDW. On the other hand, the periodic modulations observed between $-800$~mV and $+50$~mV correspond to well-defined \textbf{\textit{q}}-vectors (Figure~\ref{fig: Collected123}e) and comply with all the criteria expected for a CDW: their \textbf{\textit{q}}-vector of $q_{1/2}=\frac{1}{2}q_B$ does not depend on the imaging voltage, they appear alongside a gap in the local DOS at the Fermi level, and their contrast is inverting across this gap in topographic and spectroscopic images.\cite{spera2020} The broad $q_{1/2}$ components reflect the lack of long-range order and limited size of the CDW domains. Here, the Bragg peaks are similarly diffuse, consistent with a commensurate CDW at $q_{1/2}$ whose higher-order components will coincide with and affect the peaks at $q_{B}$.

\begin{figure}[!htbp]
\centering
  \includegraphics[width=.5\linewidth]{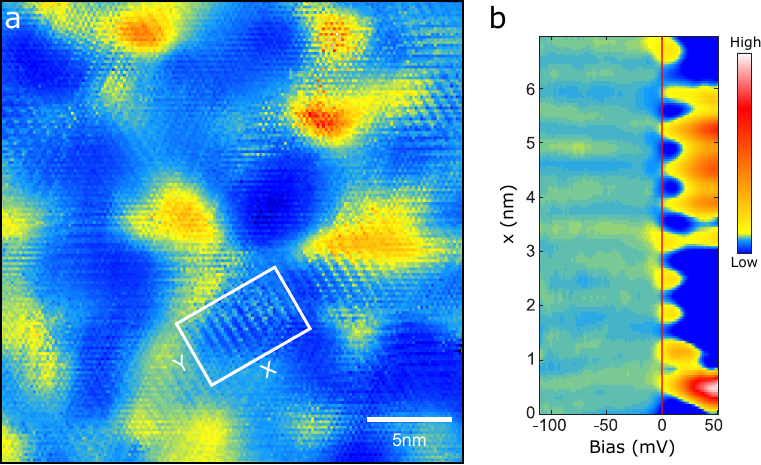}  
\caption{(a) Normalized STS map at $-9$~mV revealing CDW modulations. (b) $dI/dV(V)$ map as a function of position and energy averaged along the $Y$ direction at each position $X$ over the white rectangle in (a), showing the CDW contrast inversion across $E_F$.}
\label{fig: FigContrastCollaps}
\end{figure}

As seen in the conductance map in Figure~\ref{fig: FigContrastCollaps}a, the CDW revealed here by STM in bulk ZrSe$_2$ is a $2a\times2a$ charge modulation with different strengths of its three characteristic \textbf{\textit{q}}-vectors, giving it a 1Q, 2Q or 3Q character\cite{mcmillan1976} depending on location on the surface. The stripy character of the CDW in some regions has been reported previously on other TMD compounds. The associated suppression of some of the \textbf{\textit{q}}-vectors has been explained in terms of strain\cite{soumyanarayanan2013quantum,rahnejat2011charge,cossu2020strain} or local doping\cite{novello2017stripe}. Here, we find no clear experimental evidence for local strain in the 1Q and 2Q regions, although we cannot exclude that some of the dopant atoms might induce local strain. On the other hand, we find that the conduction band minimum is closer to $E_F$ in these regions, indicating a lower degree of electron doping compared to the 3Q regions. The CDW contrast inversion across the gap near $E_F$ is seen in Figure~\ref{fig: FigContrastCollaps}b, where we plot the conductance averaged along $Y$ as a function of position $X$ and energy in the 1Q region outlined by the white box in Figure~\ref{fig: FigContrastCollaps}a.

\begin{figure}[!htbp]
\centering
  \includegraphics[width=0.5\linewidth]{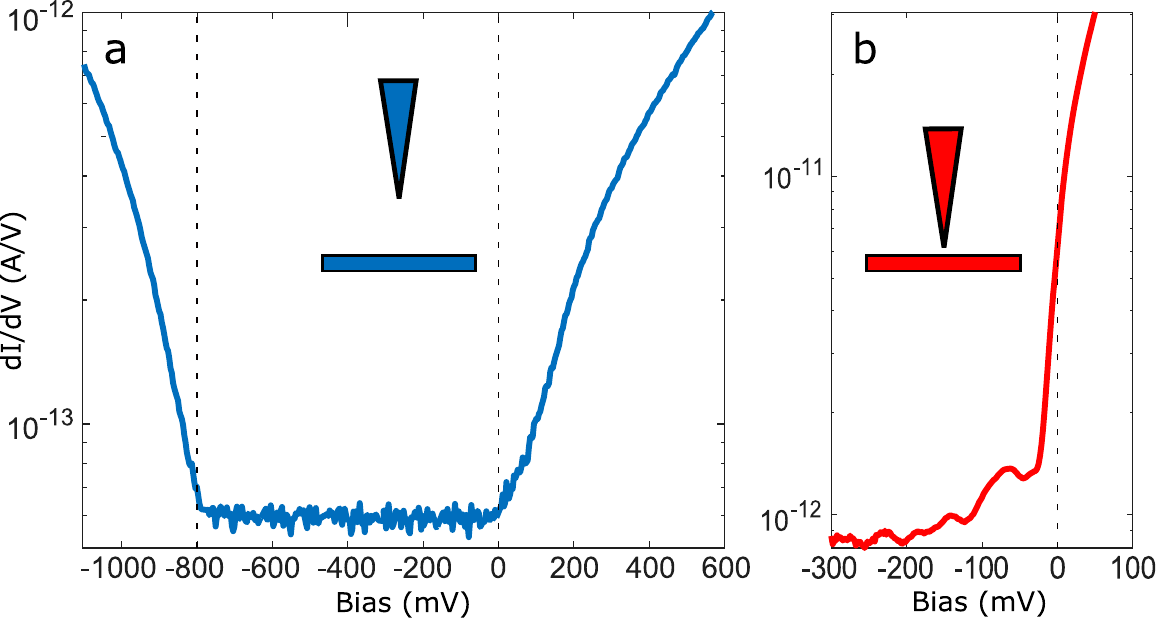}  
\caption{Two averaged $dI/dV(V)$ spectra on a log scale from a grid with different setpoint bias, $-300 $ mV and $+600 $ mV for (a) and (b), respectively. The small inserts in each graph show conceptual sketches of the corresponding tip-sample distance.}
\label{fig: STSPOSNEG}
\end{figure}

Most relevant for the present study is that the CDW primarily develops in regions where local impurities induce electron doping. This is exemplified by the strong correlation between the dark defects seen at large negative bias in Figure~\ref{fig: Collected123}c and the blue regions corresponding to finite CDW amplitudes in Figure~\ref{fig: Collected123}f. These dark defects correspond to electron doping by subsurface defects. The bright defects identified by arrows in Figure~\ref{fig: Collected123}c,f correspond to local hole doping by other subsurface defects. They shift the Fermi level below the conduction band edge, leading to a suppression of the CDW amplitude. One may realise that the contrast in Figure~\ref{fig: Collected123}b is opposite to the above defect analysis, with the CDW developing in the bright regions. This apparent contradiction is a direct consequence of the setpoint dependence of the tunnelling current in the presence of a semiconducting gap. Indeed, for a large negative bias setpoint outside the semiconducting gap shown in Figure~\ref{fig: STSPOSNEG}a, the integrated DOS available for tunnelling is reduced as the Fermi level is shifting into the conduction band (i.e. for electron doping), and the tip will have to move closer to the surface to maintain a constant tunnelling current. Hence, electron doping defects appear as depressions (i.e. dark Figure~\ref{fig: Collected123}c). The opposite is true when tunnelling at a small negative setpoint voltage within the semiconducting gap (Figure~\ref{fig: STSPOSNEG}b). In this case, the integrated DOS available for tunnelling is larger when the Fermi level is shifted higher into the conduction band. Consequently, the tip has to be retracted to maintain a constant tunnelling current and the corresponding surface region appears high (i.e. bright in Figure~\ref{fig: Collected123}b).

\begin{figure}[!htbp]
\centering
  \includegraphics[width=.7\linewidth]{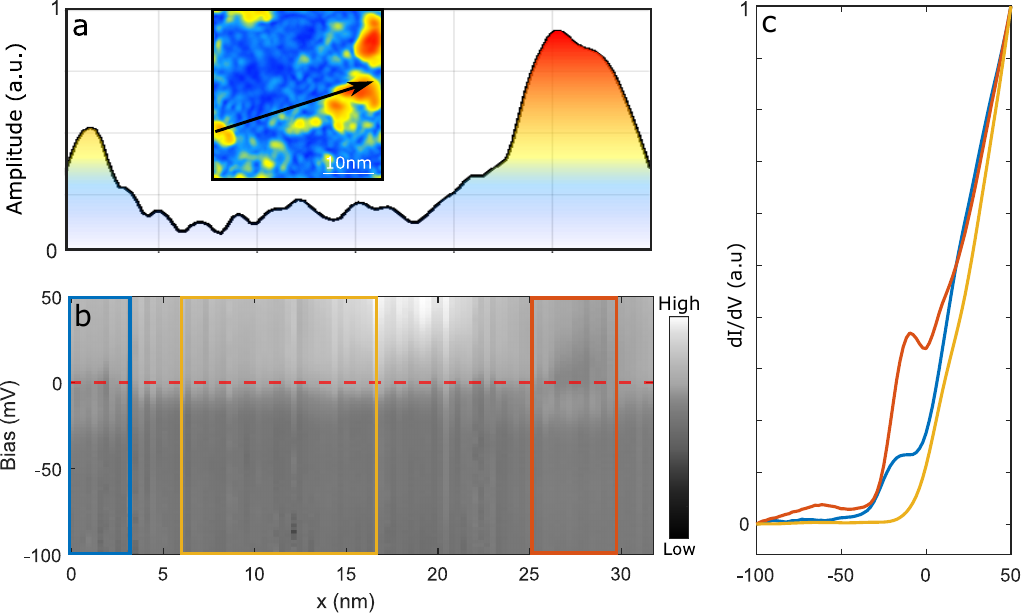}  
\caption{CDW modulation amplitude versus spectroscopic signal near $E_F$. (a) CDW modulation amplitude along the black line in the inset CITS map. (b) Grey-scale plot of the $dI/dV(V)$ spectra measured along the same line as panel (a). (c) Normalized $dI/dV(V)$ spectra averaged over the boxes with corresponding colour in panel (b).}
\label{fig: Waterfall}
\end{figure}

The direct link between electron doping and the emergence of a CDW reconstruction is very explicit in Figure~\ref{fig: Waterfall}. Figure~\ref{fig: Waterfall}a shows the amplitude of the CDW Fourier component as a function of position along the trace displayed in the inset. In Figure~\ref{fig: Waterfall}b, we present a grey-scale plot of the tunnelling conductance as a function of energy and position along the same trace. 
The minimum of the conduction band edge is at the Fermi level --highlighted by a red dashed line-- in the regions devoid of any CDW modulations.
Remarkably, we measure a finite CDW amplitude wherever the conduction band edge shifts below $E_F$, with the appearance of a slight depression in the conductance near $E_F$ corresponding to the CDW gap. To emphasise these characteristic spectral features, we consider the $dI/dV(V)$ curves in three selected boxes outlined in Figure~\ref{fig: Waterfall}b and plot their average in the corresponding colour in Figure~\ref{fig: Waterfall}c. The semiconducting gap extends to the Fermi level (yellow spectrum in Figure~\ref{fig: Waterfall}c) in the central region where no CDW modulation is detected (yellow box). Conversely, in regions where a finite CDW amplitude is detected, the edge of the conductance band is significantly below $E_F$, with a gap appearing at $E_F$ (red and blue spectra in Figure~\ref{fig: Waterfall}c). Note that the gap is more pronounced in the region outlined in red, where the CDW Fourier component is stronger compared to the region outlined in blue. 

Band structure calculations\cite{https://doi.org/10.1002/pssb.201700033} suggest that the CDW and QPI feature we observe on ZrSe$_2$ originate in different electronic bands. The CDW lives in the lowest available conduction band, whose minimum lies at the $M$--point. The next band above $E_F$ is a few hundred meV higher, with a minimum at the $\Gamma$--point. States at $\Gamma$ (i.e with small $k_{\parallel}$) decay slower into the vacuum and thus contribute more to the tunnelling current than states at the $M$-point with a large parallel momentum to the surface in a 2D crystal \cite{wiesendanger2013scanning}. Therefore, the bottom of the conduction band at the $M$--point can only be measured at low bias when the tip is close to the surface and when only the lowest energy states are sampled in the tunnelling process. By increasing the tunnelling bias, we probe deeper into the conduction band, resulting in the tunnelling current being dominated by the states at the $\Gamma$--point, which are not involved in the CDW reconstruction but contribute to QPI. This explains the different nature of the periodic modulations resolved above and below $V_\text{bias}=+100$~meV.

Angle-resolved photoemission spectroscopy (ARPES) of alkali metal and copper intercalated ZrSe$_2$\cite{muhammad2018electron,muhammad2024extrinsic,wang2019band} further support our data analysis. These experiments clearly show that electron doping from the intercalated atoms shifts $E_F$ into the conduction band and forms electron pockets at the $M$--point. The $q_{1/2}$ CDW modulation is associated with nesting vectors connecting these pockets\cite{ren2022semiconductor}. We do not introduce intentional dopant atoms into our crystals. However, stoichiometric ZrSe$_2$ single crystals are challenging to grow, and the dopant atoms in our case are most likely self-intercalated excess Zr atoms\cite{brauer1995electronic}, which are expected to electron dope the system.\cite{,wang2023first,ren2022semiconductor}. 

In summary, analysing scanning tunnelling conductance maps as a function of energy, we find that intrinsic doping induces a $2a\times 2a$ CDW modulation at the cleaved surface of bulk ZrSe$_2$. Electron doping causes the Fermi level to shift into the conduction band whose minimum sits at the $M$-point, which triggers the formation of a CDW reconstruction and the opening of a gap at the Fermi level. STS, ARPES and theoretical band structure calculations explain the set-point-specific differences in spectroscopy and topography images in terms of momentum selectivity in the STM tunnelling process. Our study unambiguously demonstrates the ability to tune the CDW phase transition by means of electron doping in a semiconducting bulk TMD. In addition to paving the way for applications exploiting tuneable CDW ground states, these results enable further studies to understand the CDW formation mechanism. Of particular interest is quasiparticle interference imaging as a function of non-local doping using field effect or space charge doping to map the band structure in the vicinity of the Fermi level in the presence or absence of a CDW over the same area. 

\begin{acknowledgement}
We thank A. Guipet and G. Manfrini for their technical assistance with the scanning probe instruments. This work was supported by the Swiss National Science Foundation (Division II Grant No. 182652).
\end{acknowledgement}

\providecommand{\latin}[1]{#1}
\makeatletter
\providecommand{\doi}
  {\begingroup\let\do\@makeother\dospecials
  \catcode`\{=1 \catcode`\}=2 \doi@aux}
\providecommand{\doi@aux}[1]{\endgroup\texttt{#1}}
\makeatother
\providecommand*\mcitethebibliography{\thebibliography}
\csname @ifundefined\endcsname{endmcitethebibliography}
  {\let\endmcitethebibliography\endthebibliography}{}

\end{document}